\documentclass[prc,aps,twocolumn,showpacs,preprintnumbers,eqsecnum,amsmath,amssymb,twocolomn]{revtex4-1}
\usepackage{graphicx}
\usepackage{epsfig}
\usepackage{dcolumn}
\usepackage{bm}
\begin{document}
\title{Pygmy resonances and nucleosynthesis}
\author{
 N.~Tsoneva$^{1,2}$, H.~Lenske$^{1}$}
\affiliation{
  $^1$Institut f\"ur Theoretische Physik, Universit\"at Gie\ss en,
  Heinrich-Buff-Ring 16, D-35392 Gie\ss en, Germany \\
$^2${Institute for Nuclear Research and Nuclear Energy, 1784 Sofia, Bulgaria}}

\begin{abstract}
  A  microscopic theoretical approach based on a self-consistent density functional theory for the nuclear ground state and  QRPA formalism extended with multi-phonon degrees of freedom for the nuclear excited states is implemented in investigations of new low-energy modes called pygmy resonances.  Advantage of the method is the unified description of low-energy multiphonon excitations, pygmy resonances and core polarization effects. This is found of crucial importance for the understanding of the fine structure of nuclear response functions at low energies.  Aspects of the precise knowledge of nuclear response functions around the neutron threshold are discussed in a connection to nucleosynthesis.
\end{abstract}
\maketitle
\section{Introduction}
\label{intro}
Recently, a new low-energy mode called pygmy dipole resonance (PDR) which reveal new aspects of the dynamics of isospin asymmetric nuclear matter has been observed \cite{Sav13}. In these findings an enhanced electric dipole strength resembling a resonance structure close to the neutron emission threshold was detected as a common feature of stable and unstable nuclei with neutron excess. It was associated with oscillations of a small outer layer of neutron-rich nuclear matter with respect to the isospin symmetric nuclear core.

\begin{figure*}
\centering
\includegraphics[width=16.5cm,clip]{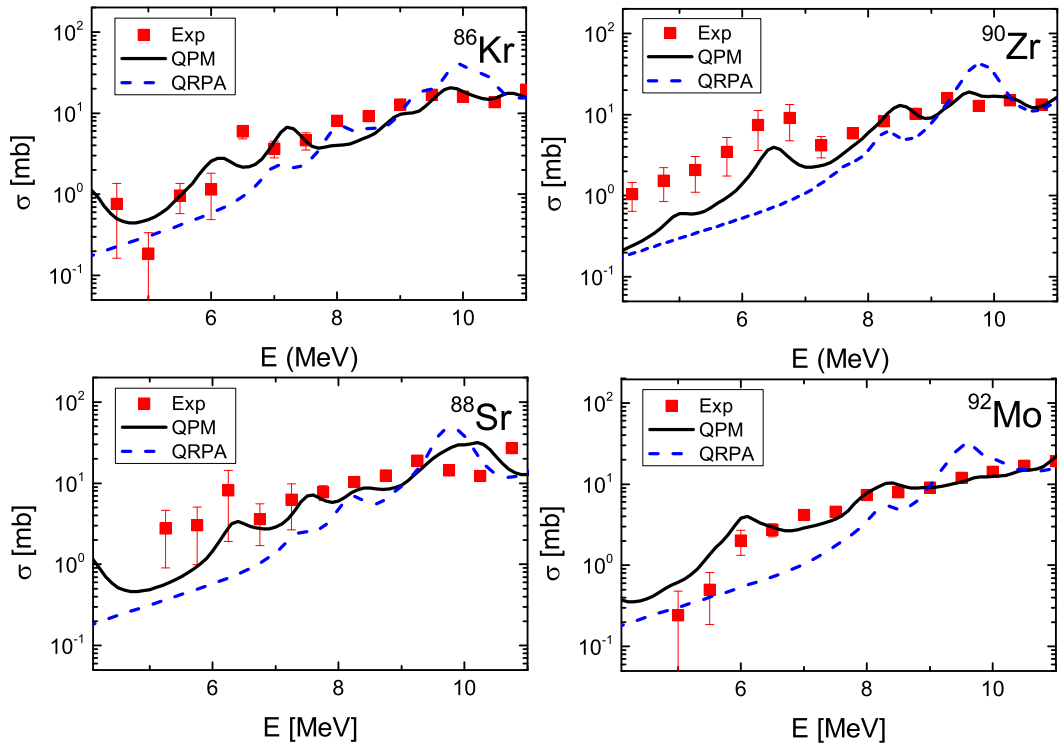}
\caption{(color online) Systematic QRPA (in blue) and three-phonon QPM (in black) calculations of dipole photoabsorption cross section below the neutron threshold of N=50 nuclei in comparison with experimental data from [6].}
\label{fig.1}       
\end{figure*}

From systematic studies of nuclear isotonic and isotopic chains a correlation of PDR strengths and nuclear skin thickness was found
\cite{Tso04,Vol06,Tso08,Schw08,Schw13}. The connection of PDR to oscillations of weakly bound nucleons at the nuclear surface was further confirmed by analysis of nuclear ground states and neutron and proton transition densities \cite{Tso08,Schw13}. 
According to recent theoretical observations the existence of PDR might have a large impact on neutron-capture reaction cross sections \cite{Gor02} contributing to the nucleosynthesis. However, those studies were based mostly on quasiparticle-random-phase-approximation (QRPA) which could not explain in details the experimental data on low-energy dipole excitations. The precise knowledge of the later is very important for calculations of nuclear reaction rates  of astrophysical importance. In this aspect, details of the fine structure and dynamics of the nuclear response function around the neutron threshold should be taken into account.
\section{The Model}
\label{sec-1}
A successful description of the pygmy resonances could be achieved in a microscopic theoretical approach which incorporates the density functional theory and  the three-phonon quasiparticle-phonon model QPM \cite{Tso04,Tso08}. Recently the method was implemented in the description of the structure of nuclear electric and magnetic excitations from the PDR region \cite{Ton10,Rus13}. \\
The model Hamiltonian is given by:
\begin{equation}
H=H_{MF}+H_{res} \quad .\label{hh}
\end{equation}
Here, $H_{MF}=H_{sp}+H_{pair}$ is the mean-field part which is obtained from self-consistent HFB theory \cite{Hofmann}. The $H_{MF}$ defines the single particle properties including potentials and pairing interactions for protons and neutrons, such that also dynamical effects beyond mean-field can be taken into account. That goal is achieved in practice by using fully microscopic HFB potentials and pairing fields as input but performing a second step variation with scaled auxiliary potentials and pairing fields readjusted in a self-consistent manner such that nuclear binding energies and other ground state properties of relevance are closely reproduced \cite{Tso08}. 

The residual interaction is based on the formalism of the QPM \cite{Sol76}:
$H_{res}$=$H_M^{ph}+H_{SM}^{ph}+H_M^{pp}$, where $H_M^{ph}$, $H_{SM}^{ph}$ and $H_M^{pp}$ are residual interactions terms
taken as a sum of isoscalar and isovector separable multipole and
spin-multipole interactions in the particle-hole (p-h) and multipole
pairing interaction in the particle-particle (p-p) channels. The model parameters are fixed either empirically \cite{Vdo} or by reference to QRPA calculation performed within the density matrix expansion (DME) of G-matrix interaction discussed in Ref. \cite{Hofmann}.

The QPM model basis is constucted of QRPA phonons defined as:
\begin{equation}
Q^{+}_{\lambda \mu i}=\frac{1}{2}{
\sum_{jj'}{ \left(\psi_{jj'}^{\lambda i}A^+_{\lambda\mu}(jj')
-\varphi_{jj'}^{\lambda i}\widetilde{A}_{\lambda\mu}(jj')
\right)}},
\label{phonon}
\end{equation}
where $j\equiv{(nljm\tau)}$ is a single-particle proton or neutron state;
${A}^+_{\lambda \mu}$ and $\widetilde{A}_{\lambda \mu}$ are
time-forward and time-backward operators, coupling 
two-quasiparticle creation or annihilation operators to a total
angular momentum $\lambda$ with projection $\mu$ by means of the
Clebsch-Gordan coefficients $C^{\lambda\mu}_{jmj'm'}=\left\langle
jmj'm'|\lambda\mu\right\rangle$.
The excitation energies of the phonons and the time-forward and time-backward amplitudes
$\psi_{j_1j_2}^{\lambda i}$ and $\varphi_{j_1j_2}^{\lambda i}$ in Eq.~(\ref{phonon}) are determined by solving the QRPA equations \cite{Sol76}.

\begin{figure}
\includegraphics[width=9cm,clip]{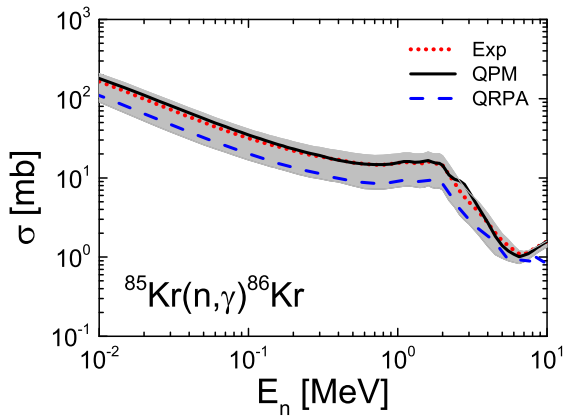}
\caption{(color online)
Cross section of $^{85}Kr^g(n,\gamma)^{86}$Kr calculated with TALYS using experimental dipole (in red), QRPA (in blue) and three-phonon QPM (in black) from Ref. [7] strength functions. The predicted uncertainties (shaded area) are derived from the experimental errors of the dipole strength function and from variations in the nuclear level density parameters.}
\label{fig.2}       
\end{figure}

Furthermore, the QPM provides a microscopic approach to multi-configuration mixing \cite{Sol76}. The wave function of an excited state consists of one-, two- and three-phonon configurations \cite{Gri94,Pon98}.

\section{Discussion}
 Systematic QRPA and QPM calculations of the electric dipole response in different isotopic and isotonic chains of nuclei \cite{Vol06,Schw08,Tso04,Tso08,Schw13} indicate enhanced $E1$ strength in the energy range below the neutron threshold with respect to  the shape of a  Lorentz-like strength function used to adjust the GDR \cite{Schw13}. 
A common observation is that the total E1 QRPA strength associated with PDR increases with the increase of the isospin asymmetry of the nucleus defined by the N/Z ratio.
Similar results are also reported by various experiments \cite{Sav13}. An exception are recent data on $^{120}Sn$ nucleus which are in a contradiction with theoretical predictions from \cite{Banu14}. This case should be further examined.

The correlation between the total PDR strength obtained in QRPA calculations and the neutron skin thickness \cite{Vol06,Schw08,Tso04,Tso08}, which in neutron-rich nuclei is defined by the differences of neutron and proton root-mean-square (RMS) radii, $\delta r=\sqrt{<r^2>_n}-\sqrt{<r^2>_p} \quad$, could be explained the following way. By definition the QRPA excited states are built only from single p-h contributions to the state vectors (see eq. \ref{phonon}). In neutron-rich nuclei, in QRPA presentation the PDR is formed by a sequence of 1$^-$ excited states, whose structure is dominated by oscillations of weakly bound, almost pure neutron two-quasiparticle configurations. The increase of the total PDR strength toward more neutron-rich nuclei could be related to the increase of the amount of those weakly bound quasiparticle neutron states around the Fermi surface which is directly connected with the decrease of the neutron binding energy and the increase of the absolute value of the difference between proton and neutron Fermi energies, $\Delta_F=\epsilon^{F}_{p}-\epsilon^{F}_{n}$ . The later is correlated linearly with the neutron skin thickness \cite{Sav13}.

QRPA calculations in N=50 isotones are shown in Fig. \ref{fig.1}. The excitation energy region below E$\leq$9 MeV is related to PDR \cite{Tso08,Schw13} whose total strength smoothly decreases with increasing proton number Z. It is closely correlated with the thickness of the neutron skin \cite{Schw13}.

As the excitation energy is increased, the isovector contribution to the dipole strength increases following closely its Lorentzian fall-off often assumed with GDR in data analyses \cite{Ber75}. Theoretically, this can be seen in transition densities and state vectors structure which manifest an enlarging of the out-of-face neutron to proton contributions and corresponding energy-weighted sum rules which is generally associated with the GDR \cite{Tso08,Schw13}.

It is clear that QRPA is unable to account for correlations of higher order like 2p-2h, 3p-3h, etc. and interactions resulting from core polarization effects. 
The later could induce dynamical effects related to redistribution of strength and strongly affect the gross and fine structure of dipole strength functions. Such effects are out of the scope of the QRPA. A comparison of QRPA calculations of dipole photoabsorption cross sections and experimental data in N=50 isotones, demonstrating the last statement, is presented in Fig. \ref{fig.1}. However, the experiment could be much better explained if the interaction between quasiparticles and phonons \cite{Sol76} is taken into account in the frame of the three-phonon QPM.
A comparison of QRPA, the three-phonon QPM and data is shown in the same figure. It indicates that for the PDR region the coupling of QRPA PDR and GDR phonons and multiphonon states is very important. The result is a shift of E1 strength toward lower energies
which can be described in three-phonon QPM formalism. Furthermore, the QPM calculations are able to reproduce in details the structure of the low energy excitations fairly well as it follows also from our previous studies of E1, E2 and M1 low-energy excitations and giant resonances \cite{Ton10,Rus13,Tso11}. Such precise knowledge of nuclear response functions is of major importance for the determination of photonuclear reactions cross sections. 

Our QRPA and three-phonon QPM microscopic strength functions have been implemented into statistical reaction code to investigate n-capture cross sections of astrophysical importance  \cite{Raut13}. As an example case, shown in Fig. \ref{fig.2}, is the calculation of the n-capture cross section of the reaction $^{85}$Kr(n,$\gamma$)$^{86}$Kr \cite{Raut13}.
It is seen that the n-capture cross section calculated with the three-phonon QPM is in a very good agreement with the experimental data \cite{Raut13} while the QRPA gives a reduced value of about $\approx$35$\%$. The estimated value of the pure PDR contribution to the QRPA calculated n-capture cross section is of the order of $\approx$30$\%$. \\
In conclusion, a common observation of our studies is that the QRPA is not sufficient to described the nuclear excitations at low energies. The analysis of experimental data confirms the importance of the QPM in astrophysical applications and in particular also the ability of the involved multiphonon theoretical methods for exploratory investigations of reaction rates in hitherto experimentally inaccessible mass regions. \\
\section{Acknowledgments}
We wish to acknowledge the support of S. Goriely in providing us with
TALYS calculations and R. Schwengner for helping us with data analysis and useful discussions. The work is supported by BMBF grant 05P12RGFTE.\\

\end{document}